\documentclass[12pt]{iopart}
\usepackage{graphicx,latexsym}
\newcommand{\p}{\partial}
\newcommand{\be}{\begin{equation}}
\newcommand{\ee}{\end{equation}}
\newcommand{\bea}{\begin{eqnarray}}
\newcommand{\eea}{\end{eqnarray}}
\newcommand{\beann}{\begin{eqnarray*}}
\newcommand{\eeann}{\end{eqnarray*}}

\newcommand{\tsh}{1/2 }
\newcommand{\bt}{\beta}
\newcommand{\g}{\eta}
\newcommand{\orow}{\bar{O}}
\newcommand{\oc}{\bar{O}^{\dagger}}
\def\bbbone{{\mathchoice {\rm 1\mskip-4mu l} {\rm 1\mskip-4mu l}
{\rm 1\mskip-4.5mu l} {\rm 1\mskip-5mu l}}}
\newcommand{\um}{\bbbone}

\newcommand{\efield}{\vec{E}}

\newcommand{\lb}{\left(}
\newcommand{\rb}{\right)}

\newcommand{\daysuffix}{\begingroup
\count0=\day
\count10 =\count0
\ifnum \count0>9 \advance\count0 by -10
\fi
\ifnum \count0>9 \advance\count0 by -10
\fi
\ifnum \count0>9 \advance\count0 by -10
\fi
\ifcase \count0  th\or st\or nd\or rd\else th
\fi
\endgroup}
\newcommand{\monthname}{%
\ifcase \month
\or January\or February\or March\or April\or May\or June%
\or July\or August\or September\or October\or November\or
December%
\fi}%

\begin{document}
\jl{1}
\title{Relativistic Aharonov-Casher Phase in Spin One
}
\author{James A Swansson\footnote{Present address: Atomic and
Molecular Physics Laboratories, Research School of Physical
Sciences
and Engineering, Australian National University, ACT, Australia,
0200.} and Bruce H J
McKellar\footnote{email: mckellar@physics.unimelb.edu.au}}
\address{School of Physics, University of Melbourne,\\
Parkville,
Victoria, Australia,  3052}
\begin{abstract}

The Aharonov-Casher (AC) phase is calculated in relativistic wave
equations of spin one.  The AC phase has previously been calculated
from the Dirac-Pauli equation using a gauge-like
technique~\cite{MK1,MK2}.  In the spin-one case, we use Kemmer theory
(a Dirac-like particle theory) to calculate the phase in a similar
manner.  However the vector formalism, the Proca theory, is more
widely known and used.  In the presence of an electromagnetic field,
the two theories are `equivalent' and may be transformed into one
another.  We adapt these transformations to show that the Kemmer
theory results apply to the Proca theory.  Then we calculate the
Aharonov-Casher phase for spin-one particles directly in the Proca
formalism.

\end{abstract}  \submitted


\maketitle


\section{Introduction}

The Aharonov-Casher phase is often linked, perhaps subconsciously,
with the Aharonov-Bohm phase.  There is a common misconception that
the AC phase is the dual of the AB phase, but there are important
physical and conceptual differences between them.

  For the latter, the absence of any classical field local to the
position of the particle makes the force-free nature of the origin of
the phase only too evident.  For the former, however, the situation is
reversed, the particle is all too clearly in the presence of the
electric field, and Aharonov and Casher were not too perspicuous about
the origins of the force-free effect.  Subsequently much debate has
revolved about this
issue~\cite{goldhaber89,azimov95,commonreferences}.

At an early stage of this debate a concise letter by
Goldhaber~\cite{goldhaber89} spelled out the differences between the
two effects.  These were, in addition to the obvious difference that
the particle is located in the field, that \begin{enumerate} \item
whereas in the AB effect the flux, although infinitely long, may be
curved arbitrarily, in the AC effect the line charge must be straight
and parallel to the magnetic moment, and \item there is an extra
degree of freedom in the AC effect, viz the spin orientation, which
adds a special interest, particularly when considered in the quantum
mechanical context.  \end{enumerate} The first observation, while
correct in saying that the force-free phase arises out of a particular
configuration of the particle and field, is not yet the full story.
It was not until much later~\cite{MK1, MK2,casella90,lee96} that it
was explicitly noted that this configuration also required that the
magnetic moment of the particle is perpendicular to the plane of its
motion.

The necessity of these conditions can be demonstrated by briefly
reviewing the relativistic derivation of the phase of He and McKellar
in Ref.~\cite{MK1} This derivation has the advantage that the
assumptions necessary for the AC phase to appear as an exact result
are made explicit, whereas the original derivation, based on the
Schr\"odinger equation used a weak field approximation which obscured
some of the assumptions.  The Dirac-Pauli equation  of a
charge zero particle with an anomalous magnetic moment in an
electromagnetic field\footnote{We consider the case where the field
$F^{\mu\nu}$ is a pure electric field, with $F^{0i} = E_i$ the only
non-vanishing components.}
\be
(i\gamma^{\mu}\partial_{\mu} + {1\over
2}\mu\sigma_{\alpha\bt}F^{\alpha\bt} -
m)\psi = 0.\label{diracpauli}
\ee
is to be transformed into the free Dirac equation for a wave-function
$\psi'$ which differs from $\psi$ by a phase: $\psi' = e^{i\chi}\psi$. 
 The Dirac equation for $\psi'$, when written in terms of $\psi$,
\be
(i\gamma\cdot\partial - m)e^{i\chi}\psi = 0,\label{dirac}
\ee
 will contain derivatives of the position dependent phase $e^{i\chi}$,
which will generate interaction terms in the equation for $\psi$.  For
these interactions to have the correct Pauli form the phase $\chi$
must be a path ordered line integral of a field linearly related to the electric
field, and must also have an appropriate Dirac matrix dependence.
For the phase, make  the anzatz
\be
\exp{i\chi} ={\cal P} \exp{i \Gamma \gamma^0 \mu \int^{\vec{x}} \vec{A}'\cdot
\vec{dr}'},
\ee
where $\Gamma$ is an appropriate element of the Clifford algebra
generated by the Dirac matrices, and $\vec{A}'$ is an effective
``vector potential'', related to the electric field, and ${\cal P}$
indicates path ordering of the integral in the phase.  Both $\Gamma$
and $\vec{A}'$ are to be determined from the condition that the
equation (\ref{dirac}) generates the equation (\ref{diracpauli}).

By eliminating the constant phase component after the operation of the
derivative the conditions for the  equivalence of (\ref{diracpauli})
and (\ref{dirac}) are then
\bea
e^{-i\chi}\gamma^\nu e^{i\chi} \p_\nu \psi & = & \gamma^\nu
\p_\nu
\psi, \label{cond1}\\
e^{-i\chi}\gamma^\nu e^{i\chi} \p_\nu \psi  & = & \mu
\sigma^{\alpha\beta} F_{\alpha\beta}.\label{cond2}
\eea

Making use of the Baker-Hausdorff formula
\be
e^{-i\lambda\xi_{3}}\,\beta^{\mu}\,e^{i\lambda\xi_{3}} =
\beta^{\mu} + {\cal P}(-i\lambda)[\xi_{3},\beta^{\mu}] + {1\over
2!}{\cal P}\lb-i\lambda\rb^2 [\xi_{3},[\xi_{3},\beta^{\mu}]] \dots\
,\label{bakerhausdorff}
\ee
the
condition (\ref{cond1}) can be expressed in terms
of  a commutation condition,
\be
\left[\gamma^\nu, \Gamma \gamma^0 \right] \p_\nu \psi = 0
\ee
 which can  be satisfied only if the operator $\Gamma$ is the product
of two spatial gamma matrices, say $\Gamma = \gamma^1\gamma^2$, {\em
and} the wave-function is independent of the  spatial co-ordinate
corresponding to the  the gamma matrix which does not appear in
this equation.  In this case, $\p_3 \psi = 0$. 
For future use we note that we may rewrite\footnote{We define
$\gamma_5 = i \gamma^0\gamma^1\gamma^2\gamma^3$. Note we will also
omit further reference to path ordering for simplicity.}
\be
i\Gamma \gamma^0 = \gamma_5\gamma^3
\ee
which is the operator which has as eigenstates those states with
well defined components of the spin in the $3$ direction, when the
 motion of the particle is normal to the $3$ direction~\cite{I-Z80}. 

The choice of the operator $\Gamma$, with equation
(\ref{dirac}),
then means that the term $\gamma^i \gamma^1 \gamma^2 \gamma^0
A'_i$
must be $\sigma^{0j}E_j$, ie
\be
A'_1 = - E_2, \quad \quad A'_2 = E_1
\ee
which may be conveniently written as
\be
A'_i = - \varepsilon_{ij} E_j
\ee
with $\varepsilon_{ij}$ the two component
antisymmetric tensor with the value
$\varepsilon_{12} = +1$.

This derivation clearly shows that the dynamics are restricted to
$2+1$ dimensions through the conditions $\partial_{\perp}\psi = 0,\
E_{\perp} = 0,\ \partial_{\perp} E_{\perp} = 0$, apply, and $\psi$ is
an eigenstate of $\sigma_{\perp}$, where subscript ${\perp}$ indicates
the coordinate perpendicular to the plane of motion~\cite{MK1}.

The wave function $\psi$ is related to the solution of the free
equation (\ref{dirac}) by a well defined, coherent phase.  Therefore
we infer that it does not alter any kinematic properties of the
solution.  The interaction so isolated can therefore be referred to as
`force-free'.  A particle with zero charge but with an anomalous
magnetic moment will accumulate a topological phase, when a strict
configuration of velocity, magnetic moment and the line of charge is
is satisfied.  This will be referred to as the AC configuration.

Because these conditions have been shown to lead to the AC phase
without approximation only for a Dirac particle, yet we know from the
original non-relativistic approach that they lead to the AC phase in
the weak field approximation for arbitrary spin, we ask whether it is
possible to extend the exact derivation to relativistic higher spin
systems.  Here we take the first step in providing an answer to that
question by showing that the derivation of reference~\cite{MK1} can
indeed be extended to the relativistic spin one particle.

\section{Aharonov-Casher phase in Spin one}
\subsection{The Kemmer Theory}\label{kemmer}

This method  exploits the genuine first order nature
of the Dirac theory, where the derivative of the path-dependent phase
produced terms linear in the field strengths, and hence comparable to
the elements of the Dirac-Pauli interaction term.  So naturally we
extend this method to spin one using the first order Kemmer spinor
theory~\cite{kemmer39}.

The $16$ dimensional spin one spinor $\phi$ has a Dirac-like free
 equation of motion,
\be
\left(i\beta^{\mu}\partial_{\mu} - m \right)\phi =
0,\label{Kemmereq}
\ee
where the $\bt$-matrices are generalizations of the Dirac gamma
matrices.  These satisfy an algebra ring, which for spin one is
\be
\bt^{\lambda}\bt^{\mu}\bt^{\nu} + \bt^{\nu} \bt^{\mu}
\bt^{\lambda} =
\g^{\lambda\mu}\bt^{\nu}+
\g^{\mu\nu}\bt^{\lambda}.\label{kemmerconditions}
\ee
These Kemmer $\bt$-matrices are reducible, that is
the $16\times 16$ representation decomposes into three separate
representations: a one dimensional trivial representation; a 5-d spin
zero
representation; and the 10-d spin one representation.  Of course
these
each satisfy (\ref{kemmerconditions}) separately.  It is also
noteworthy that this algebra ring is `odd', that is it cannot
reduce
the matrix operator to the identity, unlike the Dirac algebra.

Just as in Dirac theory, the Lorentz
invariance
of the Kemmer theory  entails a
transformation of
 the spinor so that
the
matrix representation remains the same.  The Lorentz generator
for
these transformations, $S_{\mu\nu}$,  like its spin \tsh
equivalent
$\sigma_{\mu\nu}$, is proportional to the antisymmetric product
of two
matrices of the ring:
\[
S_{\mu\nu} = b\left( \bt_{\mu}\bt_{\nu} - \bt_{\nu}\bt_{\mu}
\right)
\]
These generators satisfy well known commutation relations
(see Refs.~\cite{greiner90,umezawa56}) and define
the spin operators.  The coefficient $b$ is linked to the
coefficient in the Lorentz transformation, and hence to the
coefficient of the commutation relations, and is set below according
to our convenience.

The equation of motion of a spin one neutral particle with an
anomalous magnetic moment in Kemmer theory is  
\be
\left(i\beta^{\mu}\partial_{\mu} + {1\over 2}\mu
S_{\alpha\bt}F^{\alpha\bt} - m \right)\phi = 0.\label{spinoneeom}
\ee
The interaction term does emerge from the derivation of a `2nd order
Kemmer equation' following the method of Umezawa
\cite{umezawa56,nowakowski98}, however we rely on the reduction of
this term to its non-relativistic equivalent to confirm this form as
well as fix the ambiguity in coefficients outlined above.  In
particular this is done so that the sought-for phase can be compared
with its \tsh equivalent.  We also note in advance that this term is
transformed into its equivalent in Proca theory \cite{greiner90}.

The operator component of the phase  in the spin \tsh  AC phase
solution was
found to be a pseudo-vector spin operator.  This operator can be
re-written as the anti- symmetric product of three matrices of
the
Dirac algebra, which suggests we adopt a  spin one pseudo-vector
operator  defined by
\be
\xi_{\mu} = {i\over 2} \varepsilon_{\mu \nu\lambda \rho}
\beta^{\nu} \beta^{\lambda}
\beta^{\rho}, \label{xi}
\ee
  This is easily verified
to be
 a spin
operator
in the rest frame.

Now a  path dependent phase proportional to  $\xi_3$ is
introduced in
 the
free
Kemmer equation of  motion
(\ref{Kemmereq}),
\be
\left(i\beta^{\mu}\partial_{\mu} - m \right)
e^{i\xi_{3}\int^{r}{\vec A}'\cdot dr}\phi = 0,
\ee
with the intention to transform this into the equation of motion
(\ref{spinoneeom})  with the anomalous magnetic moment term.
Allowing
the operators to act on the solution $\phi'$, and eliminating
the
phase supplies two conditions:
\bea
e^{-i\xi_{3}\int^{r}{\vec A}'\cdot dr}\,\beta^{\mu}\,
e^{i\xi_{3}\int^{r}{\vec A}'\cdot dr} = \beta^{\mu},\ \
\mbox{and}\label{firstcond}  \\
-\beta^{\mu}\xi_{3}A'_{\mu}\phi = {1\over 2}\mu
S_{\alpha\bt}F^{\alpha\bt}\phi= \mu  S_{0l}F^{0l}\phi.
\label{secondcond}
\eea
Making use of the Baker-Hausdorff formula, the first condition reduces
to the requirement that the commutator is zero. If $\mu \neq 3$ then
this is automatically satisfied, however for $\mu = 3$, inspection of
the definition of $\xi_3$ and the $\bt$-algebra will show that the
commutator does not vanish.  Thus in order to satisfy the first
condition the dynamics of the system are restricted to $2 + 1$
dimensions, just as for spin half in Ref~\cite{MK1}.  In particular
$\partial_3\phi$ and $A'_3$ are zero.

Of the second condition, consider first the operators of the LHS.  For
$\mu = 0$ we get a solution which corresponds to $F_{12}$, which the
sake of this calculation is zero.  For $\mu = 1,2$ we have (repeated
indices are no longer summed):
\be
-\beta^{\mu}\xi_3 = \beta_{\mu}\xi_3 = -\epsilon_{\mu\nu}
{1\over
2}S_{0\nu}\beta^2_{\mu}  \left(=-\epsilon_{\mu\nu} {1\over
2}\beta^2_{\mu}S_{0\nu}\right), \label{one}
\ee
where $\epsilon_{\mu\nu}$ is an antisymmetric tensor with
$\epsilon_{12} = 1$.  Then
for $\mu = 1,2$ the action of $\beta_{\mu}$  on (\ref{one})
gives
\be
- {1\over 2}\eta_{\mu\mu} \epsilon_{\mu\nu} \beta_{\mu}S_{0\nu}
= +{1\over
2}\epsilon_{\mu\nu} \beta_{\mu}S_{0\nu}.\label{two}
\ee
Employing this, the definition of $\xi_3$ and the $\bt$-algebra,
\be
s_3 S_{0\nu}\beta_{\mu}^2 \phi_s = - S_{0\nu} s_3\phi_s,
\ee
where the states  $\phi_s$ are eigenstates of the spin operator
$\xi_3$.
Then it follows that $ S_{0\nu}\beta_{\mu}^2 \phi_s$ equals
$- S_{0\nu}
\phi_s$, hence for the operators of the LHS of
(\ref{secondcond})
acting on $\phi_s$ equal ${1\over
2}\epsilon_{\mu\nu}S_{0\nu}\phi_s$
for any value of $s_3$.   Now the comparison of these terms with
the those of the Dirac-Pauli interaction term in the RHS of
(\ref{secondcond}) yields the  field conditions:
\be
A'_1 = -2\mu E_2\ \ \ A'_2 = 2\mu E_1\ \ \ E_3 = 0
\ee
Clearly given that $E_3 = 0$, and adding the further
restriction that
$\partial_3E_3 = 0$, then the  curl of  ${\vec A}'$ is equal to
$2\mu\nabla\cdot {\vec E}$.  When  the  particle is moved around
a
closed path in this configuration the AC phase is given by,
\be
\phi_{AC} = \xi_3\oint \vec{A}'\cdot dr =
2\mu\xi_{3}\int_{S}(\nabla\cdot \efield)\cdot dS =
2\mu\xi_{3}\lambda,
\ee
where $\lambda$ is the line density of charge. The conditions for the
AC effect with spin one particles are exactly those for spin 1/2,
except that the spin operator and spinor have changed.  The factor of
two shows that the phase is twice that accumulated by a spin half
particle with the same magnetic moment coupling constant, in the same
field.

\subsection{Equivalence of AC Phase in  Proca
Theory}\label{equivalence}

For the description of spin one particles, however, the Kemmer theory
is not as widely known nor used as the Proca wave equations.  Moreover
these two theories are equivalent in the sense that, including the
electromagnetic interactions, they can be transformed into one
another~\cite{greiner90}.  Therefore we should like to demonstrate
that the results obtained above are the same as those obtainable in
Proca theory.  However in Proca theory which, being inherently a
second order theory, produces derivatives of the fields which are not
present in the electromagnetic interaction at this order, this phase
technique `fails'.  So as a first step we demonstrate this equivalence
by the transformation of the AC phase obtained in the Kemmer theory
into the Proca formalism.

The transformation from the Kemmer to the Proca formalism is
usually
introduced in the context of selecting projection operators
which pick
out the spin one irreducible representation from the
$\bt$-matrices~\cite{greiner90}.
These operators then reproduce the Proca equation.  The
projection
matrices $U^{\mu}$ and $U^{\mu\nu}$ are  constructed from the
$\bt$-matrices:
\bea
U^{\mu}&=& - (\bt^{1})^2 (\bt^{2})^2 (\bt^{3})^2
(\bt^{\mu}\bt^{0} - \g^{\mu 0}),\\
U^{\mu\nu}&=& U^{\mu}\bt^{\nu} = -U^{\nu\mu}.\\
U^{\mu} \bt^{\nu} \bt^{\sigma}&=&\delta^{\nu\sigma}U^{\mu} -
\delta^{\mu\sigma}U^{\nu}
\eea
The four vector components of the Proca theory $\psi^{\nu}$, and
six
field components $G^{\mu\nu}$ are defined respectively by the
relations
\[
U^{\nu}\phi = i\sqrt{m}\psi^{\nu},\ \ \ \ U^{\mu\nu}\phi =
{1\over
\sqrt{m}}G^{\mu\nu}.
\]
Then the operation of $U^{\mu\nu}$ on the Kemmer equation yields
the
 the anti-symmetric field tensor expressed in terms of the vector components,
\be
G^{\mu\nu}=\partial^{\mu}\psi^{\nu}-\partial^{\nu}\psi^{\mu}.\label{procafield}
\ee
while the operation of $U^{\nu}$ on the Kemmer equation yields
the
Proca equation:
\be
\partial_{\mu}G^{\mu\nu} + m^2\psi^{\nu} = 0.\label{procavector}
\ee

At this stage we merely comment if the Kemmer spinor is modified
by a
simple (c-number, path-independent) phase,  $e^{i\lambda}\phi$,
then
these transformations produce a Proca vector,
$e^{i\lambda}\psi^{\nu}$, and field tensor, $e^{i\lambda}
G^{\mu\nu}$,
similarly modified by this phase.

Earlier we indicated that the anomalous magnetic moment term which in
eq. (\ref{spinoneeom}) was expressed in terms of the Lorentz operator
$S_{\mu\nu}$ contains an ambiguity.  Therefore, in the following, it
is convenient to drop the notation of equation (\ref{spinoneeom}) and
express the anomalous magnetic moment in its `absolute' form using the
commutator of $\bt$-matrices
\[
i\mu \left[\bt^{\mu},\bt^{\nu}\right]\ F_{\mu\nu}.
\]
The transformation of this term yields $- 2im\mu\left(F_{\sigma}^{\
\mu}\psi^{\sigma}\right)$.  This can be compared with that derived
from the introduction of minimal coupling in the Proca equations or
with the standard anomalous interaction form:
\[
i\mu m\hat{I}^{\mu\nu}_{\ \ \alpha\bt} F_{\mu\nu}\psi^{\bt} =
i\mu m\left(
g^{\mu}_{\ \alpha} g^{\nu}_{\ \bt}  +g^{\mu}_{\ \bt} g^{\nu}_{\
\alpha} \right)F_{\mu\nu}\psi^{\bt}=-2i\mu m
F_{\bt\alpha}\psi^{\bt}.
\]
The factor of $m$ is introduced so that the magnetic moment $\mu$ has
the usual  dimensions
$[\mu] = -1$.

As noted above, while the spin operator
$\xi_{\mu}$ commutes with the $\bt$ matrices  for unequal
indices,
this is not true when the indices are equal.
Consequently $\xi_{\mu}$ cannot commute with the  transformation
matrices $U^{\mu}$ (and therefore $U^{\mu\nu} =
U^{\mu}\bt^{\nu}$),
because $U^{\mu}$ contains all of the $\bt$ matrices at least
once.
Therefore we cannot simply derive the equivalent free Proca
equation
with a spin dependent phase.

However in the case
that the fields are projected onto specific spin states, the
action of
the pseudo-spin operator $\xi_{3}$ on the Kemmer field $\phi$
yields
the same eigenvalue $s_3$ as the spin operator $\tilde{S}_3$
acting on
the Proca fields $\psi^{\mu}$ and $G^{\mu\nu}$ derived from that
Kemmer
spinor using the relations given below.  Therefore the
magnitude/sign
of the phase acquired in the Kemmer formalism is the same as
that in
the Proca formalism. Then  the c-number phase $e^{i\lambda
s_3}\phi$ which modifies the Kemmer spinor does, as noted above,
apply
to both the Proca vector and the field tensor, and we can then
generalise to the operator phase $e^{i\lambda\tilde{S}_3}$
without a
penalty for inexactness, under the standard conditions for
Aharonov
Casher effect.

Speaking in general terms, the Kemmer equation can be thought of
as an
operator on the Kemmer wave function
\[
O_K[{\bf A}]\phi_K = 0,
\]
and the  transformation operator $\cal T$ is defined such that
the
ten independent Proca fields $\psi_P = {\cal T}\phi_K$.   Then
for a
spin dependent phase
\be
{\cal T}e^{i\lambda\Sigma_3}\phi_{Kfree} = {\cal T}e^{i\lambda
s_3}\phi_{Kfree} = e^{i\lambda s_3}\psi_{Pfree} = e^{i\lambda
\hat{S}_3}\psi_{Pfree}.
\ee

In order to show this we choose a specific representation for
the 10 component Kemmer spinor $\phi = (\phi_1, \phi_2, \ldots,
\phi_{10})^T$, and the related
$\bt$-matrices.  Our choice of the spinor components in
terms of the Proca 4-vector $\psi^\mu$ and the Proca field strength
$G^{\mu\nu}$ is exactly that given by Greiner~\cite{greiner90},
equation (15.43).
Then the Proca equations  can be written in the Kemmer form as
\[
\left(i\bt_{\mu}\partial^{\mu} - m\right)\phi = 0
\]
where the $10 \times 10$ matrices are:
\be
\bt^0 =\pmatrix{\O&\O&\um&\bar{O}^{\dagger}\cr
\O&\O&\O&\bar{O}^{\dagger}\cr
\um&\O&\O&\bar{O}^{\dagger}\cr \bar{O}&\bar{O}&\bar{O}&0},\ \ \
\bt^k=\pmatrix{\O&\O&\O&-iK^{k\dagger}\cr
\O&\O&S^k&\bar{O}^{\dagger}\cr
\O&-S^k&\O&\bar{O}^{\dagger}\cr -iK^k&\bar{O}&\bar{O}&0}
\label{betamatrices}
\ee
where the elements are
\bea
&&\O= \pmatrix{0&0&0\cr0&0&0\cr0&0&0},\ \ \um= \pmatrix{1&0&0\cr
0&1&0\cr0&0&1},\nonumber\\
&&S^1=i\pmatrix{0&0&0\cr0&0&-1\cr0&1&0},\ \
S^2=i\pmatrix{0&0&1\cr0&0&0\cr-1&0&0},\ \
S^3=i\pmatrix{0&-1&0\cr1&0&0\cr0&0&0}  \nonumber\\
&&K^1=\pmatrix{1&0&0},\ \ K^2=\pmatrix{0&1&0},\ \
K^3=\pmatrix{0&0&1}\nonumber\\
&&\orow= \pmatrix{0&0&0}\label{matrixelements}
\eea
It should be remarked that these $\bt$ matrices differ from those
given by Greiner~\cite{greiner90}.

The spin operator for the Proca   vector  is given by:
\be
S_3 = \pmatrix{0&0&0&0\cr0&0&-i&0\cr0&i&0&0\cr0&0&0&0}.
\ee
Then from this, for $s_3 \neq 0$, $(s_3)^2 = 1$,  we deduce $\psi^1 =
-is_3\psi^2$,
$\psi^0=\psi^3 = 0$.  The tensor components dependent on these
are
$G^{01}$, $G^{02} = is_3G^{01}$, $G^{23}$, $G^{13} =
-is_3G^{23}$, and
\[
G^{12}  = -is_3\partial_{\mu}\psi^{\mu} = 0.
\]
 The condition on
the polarization 4-vector that
$s^{\mu}p_{\mu}=0$, requires, for a spin orientation in the
3-direction,
 $\partial^3\psi^{\mu}=0$.  This with the condition that  $\psi^3 =0$, gives
$G^{23}=0$.
As $S_3$ commutes with momentum operators we can express all
these
relations in the eigenfunction:
\[
\tilde{S}_3\pmatrix{G^{01}\cr is_3G^{01}\cr 0\cr G^{23}\cr
is_3G^{23}\cr 0\cr \psi^1\cr is_3\psi^1\cr 0\cr 0} =
s_3\pmatrix{G^{01}\cr is_3G^{01}\cr 0\cr G^{23}\cr
is_3G^{23}\cr 0\cr \psi^1\cr is_3\psi^1\cr 0\cr 0}
\]
with the $10\times 10$ spin matrix $\tilde{S}_3$,
\be
\tilde{S}_3=\pmatrix{S_3& \O&\O&\oc\cr \O&S_3&\O&\oc\cr
\O &\O & S_3&\oc\cr \orow&\orow&\orow&0}.
\ee
 In the Kemmer theory the spin operators are
$\Sigma_i =
i\varepsilon_{ijk}\left[\bt_j,\bt_k\right]$.  The $\Sigma_3$
matrix
is easily confirmed to be ${S}_3$, i.e. the spin matrices from
the two
theories are
identical.  What, then, of the pseudo-spin operator $\xi_3$?
Explicitly $\xi_3 = i\lb\bt^0\bt^1\bt^2 + \bt^1\bt^2\bt^0
+\bt^2\bt^0\bt^1\rb$. So  this matrix is
\be
\xi_3=\pmatrix{\O&\O&S_3&\oc\cr\O&\O&\O&-iK^{3\dagger}
\cr S_3&\O&\O&\oc\cr\orow&iK^3&\orow&0}
\ee
which does not look like $\Sigma_3$. However  the operator is
evaluated as (see Refs.~\cite{greiner90,umezawa56})
\be
\bt^0\xi_3 =\xi_3\bt^0 = \pmatrix{S_3& \O&\O&\oc\cr
\O&\O&\O&\oc\cr
\O &\O & S_3&\oc\cr \orow&\orow&\orow&0},
\ee
which is almost $\Sigma_3$.  The absent element operates (in the
Proca
case) on the components proportional to $G^{23}$, which vanish as we
saw.   The action
of
$\Sigma_3$ or $\tilde{S}_3$ on the appropriate column is then
equivalent to
that $\xi_3\bt^0\phi$.

This then demonstrates that the individual spin eigenstates of
the two
representations are transformed into one another.  That means
that the
spin operator in the phase may be temporally replaced with its
eigenvalue before performing the transformation between
representations and then replaced with its counterpart.  In this
way
the second element of the task of manifesting the transference
of
the AC phase from the Kemmer theory to the Proca formalism can
be
realized.  A spin one object acquires the same phase in either
theory
when an equivalent magnetic moment interaction term is used.
The
problem of deriving the AC phase in the Proca formalism directly
is
the subject of the next section.

\subsection{Calculation of AC Phase in Proca
Theory}\label{calcprocaphase}
%
The Aharonov Casher phase can be derived for the Proca vector
using
this phase technique, albeit  with the conditions discovered
earlier.
Again we
propose to identify the solution of the Proca equation with the
anomalous magnetic moment interaction  with a phase  modified
free
Proca field.  It was seen in \S \ref{equivalence} that the
transformation of the Kemmer spinor showed that both the Proca
vector
and the Proca   field were modified by the same phase (in spite
of the
fact that one is the  derivative of the other): $\psi'^{\nu} =
\exp
\lb i\hat{S}_3\int\vec{A'}\cdot d\vec{r}\rb  \psi^{\nu}$, and
$G'^{\mu\nu}= \exp\lb  i\hat{S}_3\int\vec{A'}\cdot d\vec{r}\rb
G^{\mu\nu}$.  Then the vector  $\psi^{\nu}$ satisfies both
\be
\partial_{\mu}G^{\mu\nu} -2i\mu m F^{\ \nu}_{\mu} \psi^{\mu} +
m^2
\psi^{\nu} = 0, \label{procaone}
\ee
with its subsidiary conditions, and
\be
\partial_{\mu}G'^{\mu\nu}  + m^2 \psi'^{\nu} =
0,\label{procatwo}
\ee
also with its subsidiary conditions.

At this stage it is convenient to define the vector ${\cal
F}^{\nu}$ as
the vector term of the magnetic moment interaction
\[
{\cal F}^{\nu} = F^{\ \nu}_{\mu} \psi^{\mu} =
\pmatrix{-E_1\psi^1-E_2\psi^2-E_3\psi^3\cr
-E_1\psi^0 -B_3\psi^2 + B_2 \psi^3\cr
-E_2\psi^0 +B_3\psi^1 - B_1 \psi^3\cr
-E_3\psi^0 -B_2\psi^1 + B_1 \psi^2\cr}.
\]
Then for spin projection $\pm 1$ eigenstates with  $B_k = E_3 =
0$ the
interaction term $-2i\mu m {\cal F}^{\nu}$ becomes
\be
-2i\mu m{\cal F}^0 = 2i\mu m\lb E_1\psi^1 + E_2\psi^2
\rb\label{procathree}
\ee
This is the expression to be derived from equation
(\ref{procatwo}). An immediate difficulty in this task is that
this
equation is inherently a second order equation. The action of
the derivative on the path-dependent phase leaves a remnant term
containing the field tensor, which is the first derivative of
the
Proca vector.  This derivative can be eliminated using a
relation
derived now from the subsidiary equations.

By contracting $\partial_{\nu}$ with  eqs. (\ref{procaone}) and
(\ref{procatwo}) respectively gives
\bea
m\partial_{\nu}\psi^{\nu} =  2i\mu  \partial_{\ \nu}{\cal
F}^{\nu},\\
0=  e^{is_3\int\vec{A'}\cdot d\vec{r}}
\left\{is_3\left(\partial_{\nu}A'_{\mu}\right) G^{\mu\nu} + m^2
\partial_{\nu}\psi^{\nu} + is_3m^2 A'_{\nu}\psi^{\nu}\right\}.
\eea
So if the vector $\psi^{\mu}$ is to be a simultaneous solution
of both
equation (\ref{procaone}) and (\ref{procatwo}) then it also
satisfies the
subsidiary conditions
\be
\partial_{\nu}\psi^{\nu} = 2i{\mu \over  m} \partial_{\nu}{\cal
F}^{\nu}= -i{s_3\over m^2}\left\{\left(\partial_{\nu}A'_{\mu}
\right)
G^{\mu\nu} + m^2 A'_{\nu}\psi^{\nu} \right\}.
\ee
Now if the fields $A'_{\mu}$ satisfy the conditions $A'_0 =
\partial_0
A'_{\mu} = 0$, then the first term of the second condition
vanishes,
and we have the useful relation
\be
2 i\mu  \partial_{\nu}{\cal F}^{\nu} = -is_3 m
A'_{\nu}\psi^{\nu}.\label{subsidiarylink}
\ee
Now  the expansion of equation (\ref{procatwo}) as seen above
leaves us attempting
to make the remaining term $is_3A'_{\mu}G^{\mu\nu}$ look the
part of
the anomalous magnetic moment interaction (\ref{procathree}).
At this point it is assumed  that the phase fields $A'_k$ are
the same as
those of \S \ref{kemmer}.
It is also assumed that $\psi^{\mu}$ is a spin eigenstate with
$(s_3)^2 = 1$, so that $\psi^1 = -is_3 \psi^2$, etc..  Then, by
substitution of the electric and Proca field identities:
\be
is_3 A'_{\mu} G^{\mu\nu} = 2{\mu  m} \partial^0 {\cal F}^0.
\ee

 We can take advantage of the fact that ${\cal F}^i = 0$, to replace
$\partial^0 {\cal F}^0$ with $\partial_\mu {\cal F}^\mu$, and then
introduce the identity (\ref{subsidiarylink}).  This gives  $2{\mu
m}\partial^0 {\cal F}^0 = 2{\mu m} \partial_{\mu} {\cal F}^{\mu} =
-s_3m A'_{\mu}\psi^{\mu}$.  Now the last term gives us (repeating the
substitution of the electric and Proca field identities),
\be
s_3m\left(A'^1\psi^1 + A'^2\psi^2 \right) =2i\mu \left[E_1\psi^1
+
E_2\psi^2\right]
\ee
as the zeroth element of our vector.  This is the interaction
term (\ref{procathree}) which was sought.   Naturally the
integral of
the path-dependent phase $\hat{S}_3\oint \vec{A'}d\vec{r}$
equals the
same phase as the Kemmer calculation, as that is how we have
chose
the field relations.

Hence the AC phase can be successfully calculated by
the gauge method in the tensor formalism, although with  the
assistance of the relations first derived from the spinor
theory, and
is seen to be consistent between the two spin one theories.

\section{Remarks}
The topological nature of the Aharonov-Casher effect is delicate and
its appreciation requires a subtle understanding.  These derivations
of the topological phase  from spin \tsh and spin one relativistic wave
equations illuminate some of the difficulties at the  intersection of  quantum
mechanics and classical electromagnetic theory, and those associated
specifically with the AC effect.  

Foremost are the
explicit constraints on the geometrical alignment between the electric
field, the particle momentum and the spin quantization axis.  That these
constraints are obscured in the non-relativistic derivation of the
AC effect is evident by their protracted revelation in the
literature.  In contrast they evolve out of the relativistic
derivation in an obvious manner because of the need to reduce the
number of spinor components from which can contribute to a geometric,
scalar phase.  The implications of such  strict limitations on the interpretation of
experiments which have sought to measure the Aharonov-Casher phase are
self-evident.  

Furthermore these studies indicate the 
extraordinary care which must be taken in applying semi-classical models to
spin or the magnetic moment, or even the particle's trajectory.   The
natural determination of a spinor's properties out of the
transformation properties of the SL(2,C) group contrasts with the
difficult interpolation of spinor behaviour in a non-relativistic
framework. In the  case of the AC effect, satisfaction of the
geometrical constraints for a geometric phase is dependent on the
choice of interpretation of the classical vector which substitutes for
the spin dynamics.  In many cases a phase can still be derived from
classical origins which is mistakenly thought to be the geometric
phase.  Similarly the problem of the `hidden momentum' of a magnetic
dipole is clarified by comparison with the non-relativistic equations
derived from the complete relativistic wave equations. 

Finally the existence of the AC phase for all the spin components of a
spin 1 particle, together with the geometrical constraints, mark out
the differences between the Aharonov-Bohm and Aharonov-Casher
effects, and underscore the limitations of `equivalence' methods of
studying the AC effect using the former.  Equivalence methods can
only map two possible spin states onto the charge states of the
Aharonov-Bohm effect, and consequently omit part of the richness of
the AC effect.
\section{Conclusion}
The  Aharonov-Casher phase has been successfully calculated for spin
one relativistic particles in both relativistic theories with
equivalent electromagnetic interactions.  In the Kemmer theory, the
definition of a pseudo-vector spin operator $\xi_{\mu}$ has permitted
the imitation of the gauge technique used by He and McKellar in spin
\tsh to find the AC phase, subject to the conditions familiar from
spin $1/2$.  In order to demonstrate that these results hold
in the more familiar Proca wave theory, the two equations, one with the
interaction, the other with the modified phase, have been shown to
transform into their equivalents using the established projection
operators.  Finally, with the conditions derived from the previous
studies, the Aharonov-Casher phase has been calculated directly in the Proca
formalism itself.  The coherence of these results demonstrates the
validity of the spin dependence in the AC effect for higher spins.
\ack
The authors wish to thank  Xiao-Gang He for useful discussions. One
author (JAS)
would also like to acknowledge the support of an Australian
Postgraduate Award.


\section*{References}
\begin{thebibliography}{12}
\small
\bibitem{MK1}  Xiao-Gang He \& Bruce H.J. McKellar, Phys. Lett. {\bf B256},
250, (1991).
\bibitem{MK2} Xiao-Gang He \& Bruce H.J. McKellar, Phys. Lett. {\bf  B264},
129, (1991).
\bibitem{goldhaber89}  A.S. Goldhaber, Phys. Rev. Lett., {\bf 62},  482, (1989).
%
\bibitem{azimov95} Y.I.Azimov, \& R.M. Ryndin,
JETP {\bf  61}, 453, (1995).
\bibitem{commonreferences} Y. Aharonov, P.  Pearle, \& L. Vaidman,
Phys. Rev. A, {\bf 37},  4052, (1988); S.M Al-Jaber,
Nuovo-Cimento-B. {\bf 110B},  1003, (1995);   J. Anandan,
Phys. Lett. {\bf A138}, 347, (1989);   T.H. Boyer, Phys. Rev. A, {\bf 36}, 5083,
(1987);  G.R. Freeman, \& N.H. March,  Euro. J. Phys., {\bf 
18}, 290,  (1997);  Y.D. Han \& I.G. Koh,  Phys. Lett. {\bf A167},
341, (1992); J.Q. Liang,\&  X.X. Ding, Phys. Lett. A, {\bf 
176}, 165,  (1993);  H. Rubio, J.M. Getino, \& O. Rojo, Nuovo Cimento
B. {\bf 106B},  407, (1991); K. Sangster, E.A. Hinds, S.M. Barnett, E. Riis, \&
A.G. Sinclair, Phys. Rev. A, {\bf 51}, 1776, (1995);G.  Spavieri,  Found. Phys. Lett. {\bf 3},  291,
(1990);   G. Spavieri, \& G. Cavalleri,
Europhys. Lett.. {\bf 18}, 301,  (1992); G. Spavieri,
Nuovo-Cimento-B. {\bf 109B},  45, (1994);   L. Vaidman, Am. J. Phys.,
{\bf 58}, 978, (1990);  A Zeilinger, R Gahler \& M.A. Horne, Phys. Lett. {\bf
A154}, 93, (1991).
%
\bibitem{casella90}  R.C. Casella, Phys. Rev. Lett. {\bf 65},  2217, (1990).
\bibitem{lee96}  T.-Y. Lee, Mod. Phys. Lett. B, {\bf 10},  795, (1996).
\bibitem{I-Z80} C. Itzykson and J-B Zuber, ``Quantum Field 
Theory'', McGraw-Hill, New York, 1980.
\bibitem{kemmer39}  This spin one object in SL(2,c) was first
investigated by  R.J. Duffin, Phys. Rev. {\bf 54}, 1114, (1938),
N. Kemmer, Proc. R. Soc.{\bf  173}, 91, (1939), and G. Petiau, PhD
thesis, Paris, 1936. Periodic interest in the formalism continued.  Further
descriptions of  Kemmer theory can be found in: \cite{umezawa56}; E. Fischbach, M.M. Nieto \& C.K. Scott,
J. Math. Phys., {\bf 14}, 1760, (1973); B. Vijayalakshmi, M Satharaman \&
P.M. Mathews, J. Phys. A: Math. \& Gen., {\bf 12}, 665, (1979);
\cite{greiner90}; an application to cross-section calcultions can be
seen in L. Knurth Kerr, B.C. Clark, S. Hama, L. Ray, \& G.W. Hoffmann,
Prog. Theo. Phys., {\bf 103}, 321, (2000).
\bibitem{greiner90}  W. Greiner, ``Relativistic Quantum Mechanics'',
Springer-Verlag, New York, 1990.
\bibitem{umezawa56} H Umezawa, ``Quantum
Field Theory'', North-Holland, 1956.
\bibitem{nowakowski98} M. Nowakowski, Phys. Lett., {\bf A 244}, 329, (1998).
%
%
%
%

\end{thebibliography}
\end{document}